\documentclass[%
 reprint,
superscriptaddress,
 amsmath,amssymb,
 aps
]{revtex4-2}

\usepackage{graphicx}% Include figure files
\usepackage{dcolumn}% Align table columns on decimal point
\usepackage{bm}% bold math
\begin{document}

\preprint{APS/123-QED}

\title{Mapping charge capture and acceleration in a plasma wakefield of a proton bunch using variable emittance electron beam injection}% Force line breaks with \\

\author{E. Granados}
\email{eduardo.granados@cern.ch}
 \affiliation{CERN, Geneva, Switzerland}%
\author{L. Verra}%
 \affiliation{CERN, Geneva, Switzerland}%
 \affiliation{Max Planck Institute for Physics, 80805 Munich, Germany}
\author{A.-M. Bachmann}
 \affiliation{CERN, Geneva, Switzerland}%
 \affiliation{Max Planck Institute for Physics, 80805 Munich, Germany}
\author{E. Chevallay}%
 \affiliation{CERN, Geneva, Switzerland}%
\author{S. Doebert}%
 \affiliation{CERN, Geneva, Switzerland}% 
\author{V. Fedosseev}%
 \affiliation{CERN, Geneva, Switzerland}%
\author{F. Friebel}%
 \affiliation{CERN, Geneva, Switzerland}%  
\author{S. Gessner}%
 \affiliation{CERN, Geneva, Switzerland}%  
\author{E. Gschwendtner}%
 \affiliation{CERN, Geneva, Switzerland}%
\author{S. Y. Kim}%
 \affiliation{CERN, Geneva, Switzerland}% 
 \affiliation{Department of Physics, UNIST Ulsan, Republic of Korea}% 
 \affiliation{Argonne National Laboratory, Lemont, IL, USA}% 
\author{S. Mazzoni}%
 \affiliation{CERN, Geneva, Switzerland}%  
\author{J. T. Moody}%
 \affiliation{CERN, Geneva, Switzerland}% 
\affiliation{Max Planck Institute for Physics, 80805 Munich, Germany}
\author{M. Turner}%
 \affiliation{CERN, Geneva, Switzerland}%  

\collaboration{AWAKE Collaboration}%\noaffiliation

\date{\today}% It is always \today, today,

\begin{abstract}
In the Phase 2 of the AWAKE first experimental run (from May to November 2018), an electron beam was used to probe and test proton-driven wakefield acceleration in a rubidium plasma column. In this work, we analyze the overall charge capture and shot-to-shot reproducibility of the proton-driven plasma wakefield accelerator with various UV illumination and electron bunch injection parameters. The witness electron bunches were produced using an RF-gun equipped with a Cs\textsubscript{2}Te photocathode illuminated by a tailorable ultrafast ultraviolet (UV) laser pulse. The construction of the UV beam optical system enabled appropriate transverse beam shaping and control of its pulse duration, size, and position on the photocathode, as well as time delay with respect to the ionizing laser pulse that seeds the plasma wakefields in the proton bunches. Variable photocathode illumination provided the required flexibility to produce electron bunches with variable charge, emittance, and injection trajectory into the plasma column. We demonstrate charge capture rates exceeding 15\% ($\sim$40~pC of GeV accelerated charge for a 385 pC injected electron bunch) under optimized electron injection conditions.
\end{abstract}

\maketitle

%\tableofcontents

\section{Introduction}

In the AWAKE experiment, an electron bunch is used to probe the proton-driven wakefield acceleration in plasma. During the initial experimental run in 2018, the first demonstration of electron beam acceleration was successfully achieved \cite{Adli2018}. The experiments confirmed that the injected 19~MeV electron beam was in a 10~m long plasma cell, with a maximum energy gain of up to 2.0~GeV. The shot-to-shot performance of this accelerator was still not comparable to standard linac-based electron accelerators. With a view on improving parameters such as charge capture rate and overall acceleration efficiency, the AWAKE Run 2 experiment is currently being prepared. The goal is to achieve a charge capture efficiency and an energy gain of over 90\%, while producing 10~GeV high charge electron beams \cite{2018arXiv181208550C}. 

To accomplish this goal the electron beam parameters at the plasma entrance must be accurately controlled, including: the pointing jitter, size, charge and emittance of the electron bunches. With the aim of understanding the sensitivity and limitations of AWAKE’s experimental setup, we study here the influence of the several of those parameters on the overall charge capture and reliability measured during the first experimental run.

\begin{figure}[b]
\includegraphics[width=\linewidth]{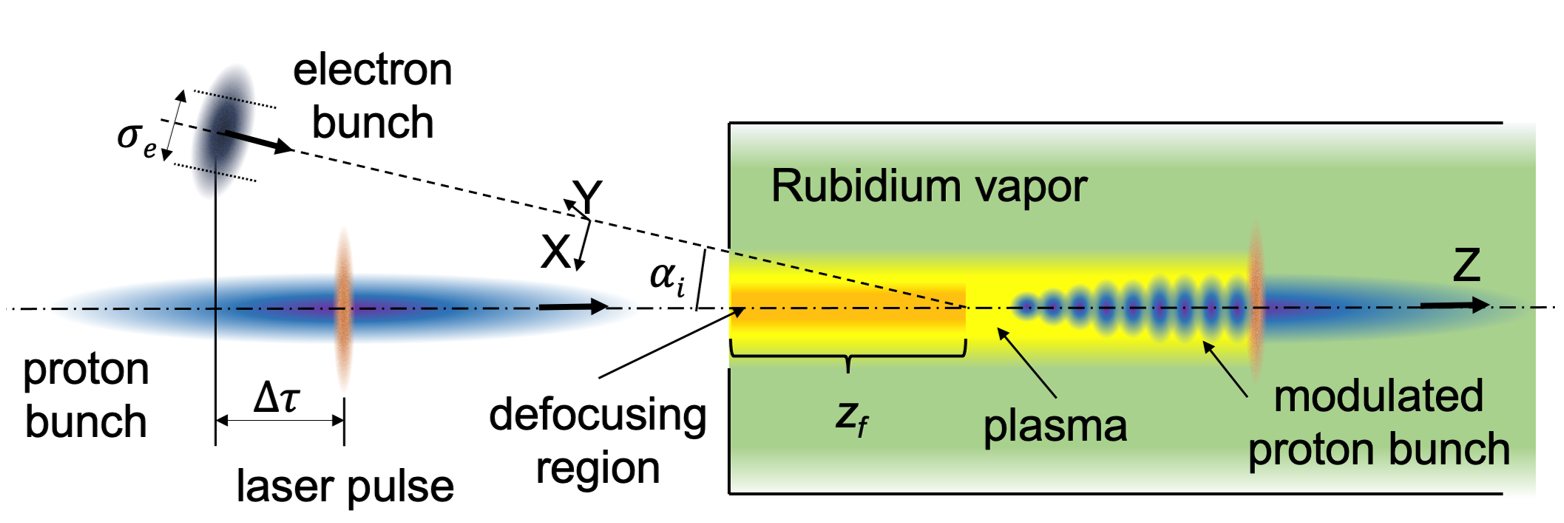}% Here is how to import EPS art
\caption{\label{fig:1} The oblique electron injection arrangement: an electron bunch of variable charge and size $\sigma_e$ is injected into the plasma wakefields at an angle $\alpha_i$}.
\end{figure}

A 400~GeV proton beam is extracted from the CERN Super Proton Synchrotron (SPS) and utilized as a drive beam for wakefields in a plasma column to accelerate electrons. The plasma is generated in a 10 m long rubidium vapour source via the over-the-barrier ionization employing a high intensity laser field. In this scheme, an ultrafast infrared laser pulse ($\sim$ 100~fs) co-propagates with the SPS proton beam and initiates a self-modulation process at the front of the plasma column. As a consequence, the long SPS proton bunch ($\sigma_z$~=~12~cm) is heavily modulated into a train of longitudinal micro-bunches which in turn drive a periodic wakefield \cite{PhysRevLett.122.054801,PhysRevLett.122.054802}, as shown in Fig. \ref{fig:1}. 

The injection of electrons into the wakefield is carried out at an angle and a distance $z_f$ with respect to the plasma entrance. This is done to avoid the loss of electrons at the density transition region at the entrance of the plasma cell (defocusing region in Fig. \ref{fig:1}). In this arrangement, the electrons approach the central axis in the region of constant plasma density and therefore get trapped into the established plasma wave.

The required injection angle $\alpha_i$ and radial offset of the electron beam at the orifice placed at the entrance of the plasma cell are small enough so the oblique injection does not require any hardware changes in the facility design, as compared to the on-axis injection. According to early theoretical models, the parameter space for good trapping is quite large when compared to the electron beam portrait \cite{CALDWELL20163,PhysRevAccelBeams.24.011301}, so no fine-tuning of the injection angle or focus point is required for the best performance.

In this paper, we further present experimental results regarding the electron acceleration performance by varying some of the main electron beam parameters, including position, angle, size, emittance and charge. The goal is not to identify a single, optimal parameter set for the witness bunch, but rather to quantify approximately the impact of these various parameters on the reliability of the acceleration, providing thus feedback for future design decisions. 

We also show experimentally that under favorable circumstances, the efficiency of the charge capture of the wakefields can easily surpass 10\%. The results are accompanied by measurements of electron beam emittance and extracted bunch charge in relation to the photoinjector UV beam parameters in the experimental parameter space during the AWAKE first experimental run.

\section{Variable emittance electron beam production}

The electron bunch is supplied by the electron beam photoinjector consisting of an RF-gun and a booster structure \cite{KIM2020163194}. The electron bunch in the RF-gun is produced using a photoemission driven by an UV beam generated from the AWAKE main laser system. The AWAKE laser installation is located in the underground TSG40 tunnel which was refurbished and equipped according to requirements for clean laser rooms \cite{Fedosseev:IPAC2019-THPGW054}. The laser system comprises a mode-locked fibre laser oscillator locked to a RF reference, a pulse stretcher, a series or Ti:Sapphire multi-pass amplifiers, and two separate grating pulse compressors (one dedicated to the ionizing beam, and a second one for generating the UV pulses for electron beam production). 

The output of the main amplifier consists of two IR laser beams with central wavelength of 780~nm. The main (primary) beam is injected into vacuum system where pulses with energy up to 660~mJ are compressed to 120~fs and transported further (450~mJ maximum pulse energy) to the rubidium vapour source for producing the plasma. The secondary IR beam with 2~mJ pulses at the repetition rate of 10~Hz is perfectly synchronized with pulses of the main laser beam since it originates from the same oscillator. The secondary beam was used to produce pulsed UV light required to generate the electron bunches. It was produced by third harmonic generation (THG) of the compressed IR pulse, producing UV pulses with energy up to 10~uJ. It was possible to control the UV pulse duration by changing the distance between diffraction gratings in the secondary beam pulse compressor. For the current experiments the compressor grating was set to a position corresponding to UV pulse duration of 5.2~ps FWHM, compared to the Fourier limited pulse duration of approximately 85~fs. The stretched UV pulse duration accounted for the relatively low conversion efficiency in the harmonic generation process. 

At the exit of the THG setup the beam was expanded using two positive lenses with the focal lengths of 100~mm and 250~mm respectively. The distance between these lenses was adjusted for producing a diffraction limited focus at a motorized aperture installed on the optical table near the electron gun. The main path connecting the THG area with the electron gun was arranged inside a straight vacuum pipe of 14.2~m length. This UV beam transfer system was designed to ensure a maximal stability of the beam and conformity to laser safety requirements. Application of this scheme was essential for minimizing the pointing instability of the UV beam on the photocathode.

The variable aperture was then imaged onto the photocathode plane using a combination of reducing telescope (M~=~1:2) and a single lens with the focal length of 1000 mm, producing a final beam on cathode with a size $\sigma$\textsubscript{UV}. The so-called “virtual cathode” setup was assembled on a small optical table installed near the electron gun. The virtual image of the beam arriving at the photocathode was produced by means of a pick-off of the UV beam and imaged onto a CMOS digital camera using a scintillating screen. 

\begin{figure}[ht]
\includegraphics[width=\linewidth]{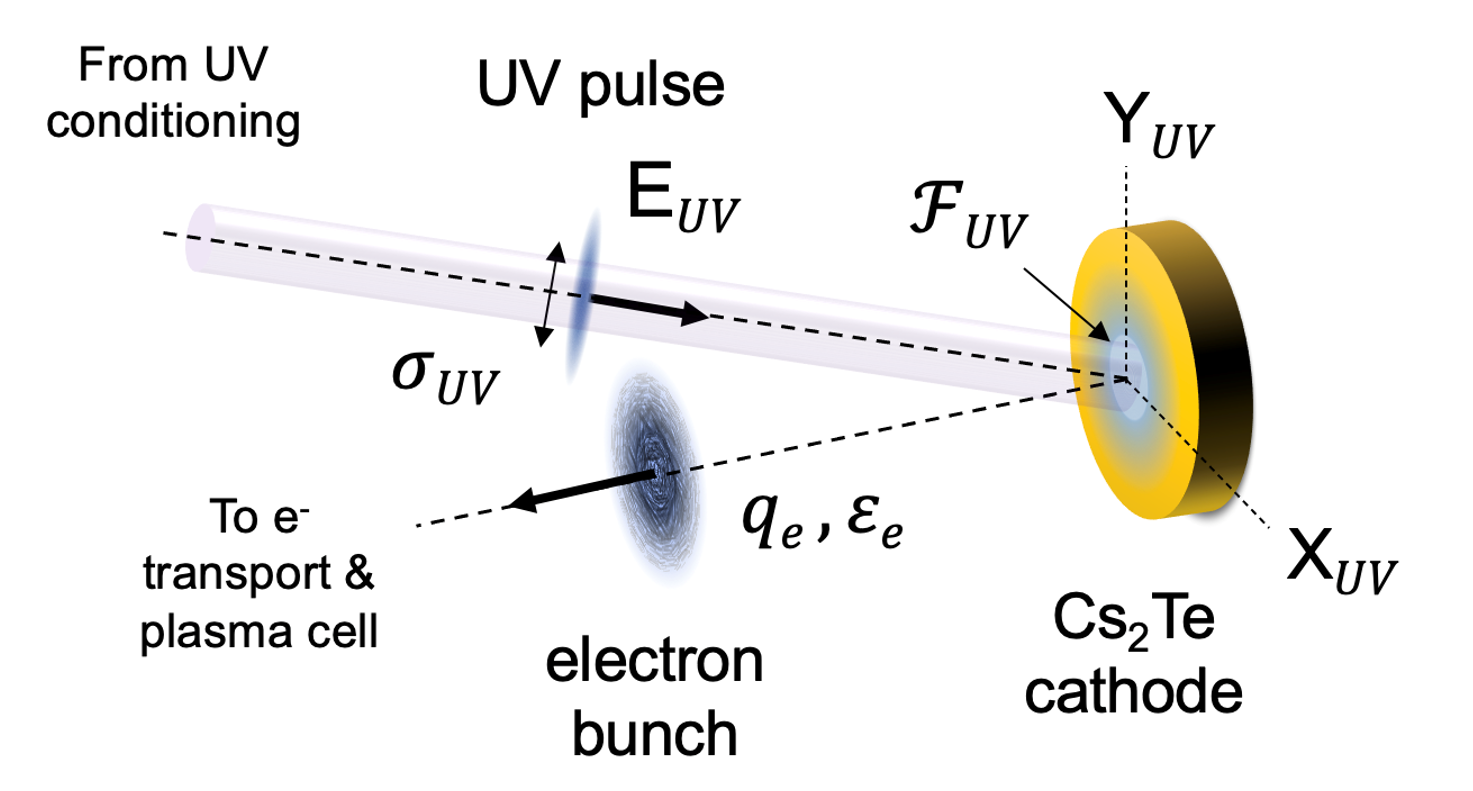}% Here is how to import EPS art
\caption{\label{fig:2} Schematic depiction of the main parameters adjusted during the experiments. The electron bunch charge ($q_e$) and emittance ($\epsilon_e$) were selected by varying the UV pulse energy (E\textsubscript{UV}) and spot size ($\sigma_{UV}$) impinging on the photocathode.}.
\end{figure}

Steering of the UV beam on the photocathode is performed using a motorized mirror mount, being possible to adjust the beam position $X_{UV}$ and $Y_{UV}$ with high precision (10s of microns). A motorized filter wheel equipped with a set of neutral density filters is used for varying the energy of UV laser pulses (E\textsubscript{UV}), while the laser energy meter provides reading of the pulse energy. The length of the UV optical path was adjusted to match arriving of electron bunches to the plasma source with ionizing laser pulses. Fine tuning of the delay time within 1~ns range is performed by means of two mirrors installed on a remotely controlled motorized stage.

The motorized aperture enables remote control of the laser beam spot size on the cathode, which in combination with control of the pulse energy delivered to the cathode, allows to vary the illumination fluences ($F_{UV}$) which in turn varies the electron bunch charge ($q_e$) and emittance ($\epsilon_e$). A summary of this arrangement is shown in Fig. \ref{fig:2}. The RF-gun was equipped with a highly efficient Cs\textsubscript{2}Te photocathode that was produced in the CERN photoemission laboratory. In particular, the photocathode used in AWAKE showed the quantum efficiency QE~$\sim$~20\% measured in the DC-gun just after the fabrication process. This cathode supplied electron beams since the commissioning in November 2017 till the end of physics runs in December 2018. After replacement by a new cathode it was analysed again in the DC-gun, showing an overall drop to QE~$\sim$~2\% allowed generation of reliable electron beams till the last days of the first AWAKE experimental run. 

The produced electron bunch length was measured using a streak camera. The optical transition radiation (OTR) light produced by electron bunches on a SiAg screen installed 2.6~m upstream the plasma source was optically transported to a streak camera, where a time profile with average FWHM duration of $\Delta \tau_e~\sim$~4~ps was measured, roughly matching the UV pulse duration.

\begin{figure}[h]
\includegraphics[width=\linewidth]{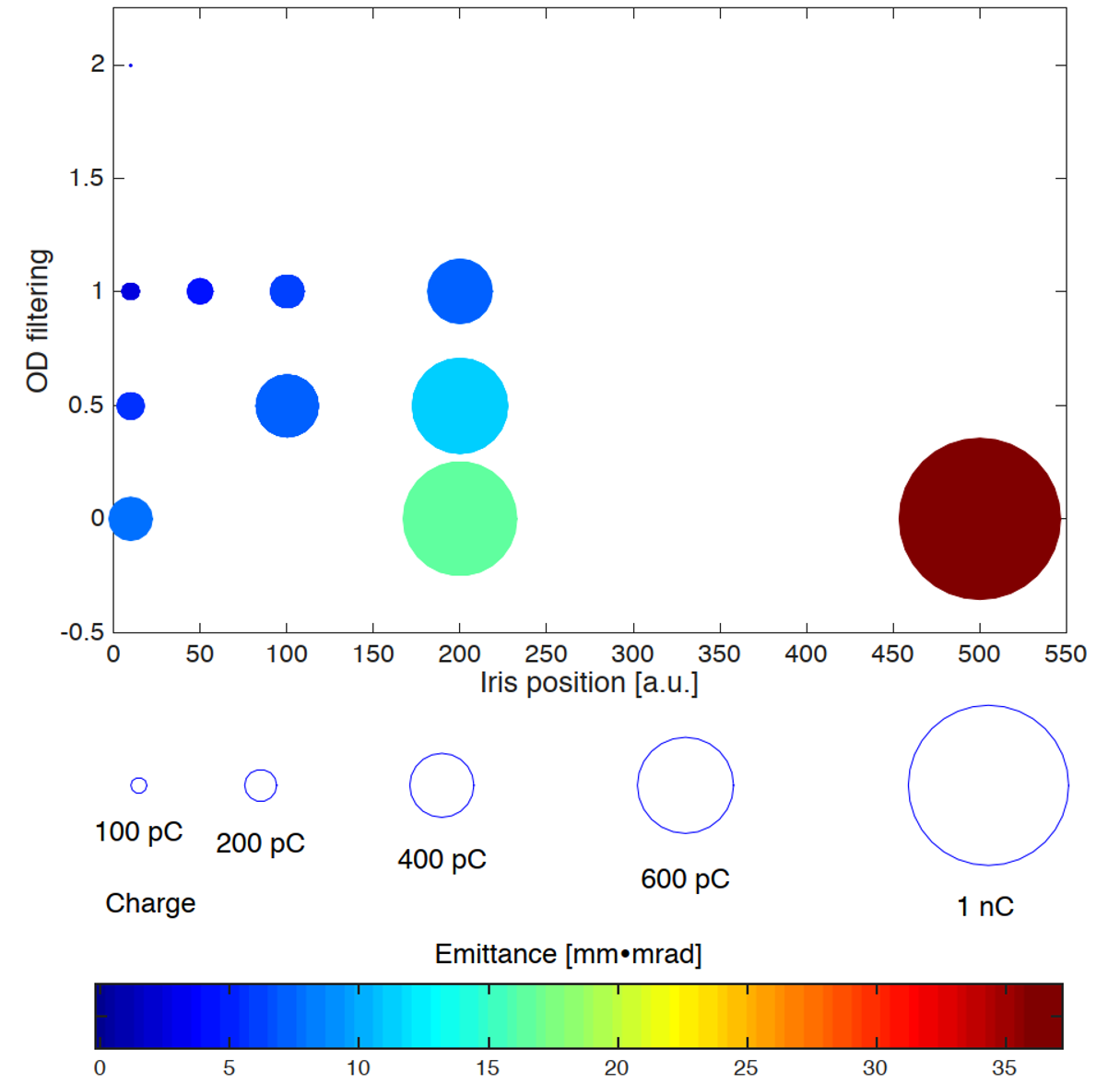}% Here is how to import EPS art
\caption{\label{fig:3} Electron beam charge and emittance for various UV photocathode illumination configurations by adjusting OD filtering and Iris positions (corresponding to UV pulse energy and spot size on cathode).}.
\end{figure}

The performance in terms of emittance and charge of the resulting electron beam is summarized in Fig. \ref{fig:3}. Here the iris position corresponds to the size of the aperture used, which modulates the exact size of the UV beam on cathode. The UV pulse energy is adjusted by employing a range of neutral density filters with various ODs. With this setup it was possible to vary the charge of the electron beam from 100~pC up to approximately 1~nC (although this measurement was limited by the Faraday cup range capabilities). In terms of emittance, the different combinations of spot size and UV energy (corresponding to the UV fluence) allowed the production of bunches with normalized emittances from less than 2~mm·mrad to up to 37~mm·mrad. Note that the electron beam size at the injection point in the plasma cell varies accordingly. These emittance was measured using RF-gun solenoid scans and pepper-pot measurements.

\section{Electron injection performance}

The RF-gun accelerates then the electron bunches to an energy of 5.5~MeV, and the booster can add a maximum energy of 16~MeV. A single klystron was used to power the gun and the booster structure. A high power waveguide attenuator and phase-shifter allow for individual phasing and powering of the two structures. More details of the electron beamline performance can be found in \cite{KIM2020163194}. 

The core of the experiment is a 10-m-long rubidium vapor source \cite{Adli2018}: a long, fluid-heated heat exchanger evaporates rubidium at 180 °C – 230 °C to reach the required vapor density of 0.5~–~10~$\times$~10\textsuperscript{14} atoms/cm\textsuperscript{3}. We used beam position monitors (BPMs) to measure the position of the proton and electron beams along the beam line and scintillating screens (BTVs) to measure their transverse bunch profiles \cite{Mazzoni:IPAC2017-MOPAB119}. Losses and radiation produced by the proton beam are monitored by proton beam loss monitors (PBLMs) positioned along the transfer line and the vapor source. 

To inject and accelerate the electrons, we spatially and temporally overlap them with the plasma wakefields \cite{8659402}. This means that the electron and proton beam trajectories have to cross within the plasma cylinder. To investigate the acceleration process and to characterize the wakefields, we injected the electron bunches at various locations of the plasma entrance and at various angles with respect to the proton beam trajectory. We observed that the highest capture and acceleration efficiency occurred when the electron beam was injected $\sim$1~m downstream from the entrance \cite{PhysRevAccelBeams.23.032803}. This was therefore the baseline setup for the acceleration experiment.

\begin{figure}[h]
\includegraphics[width=\linewidth]{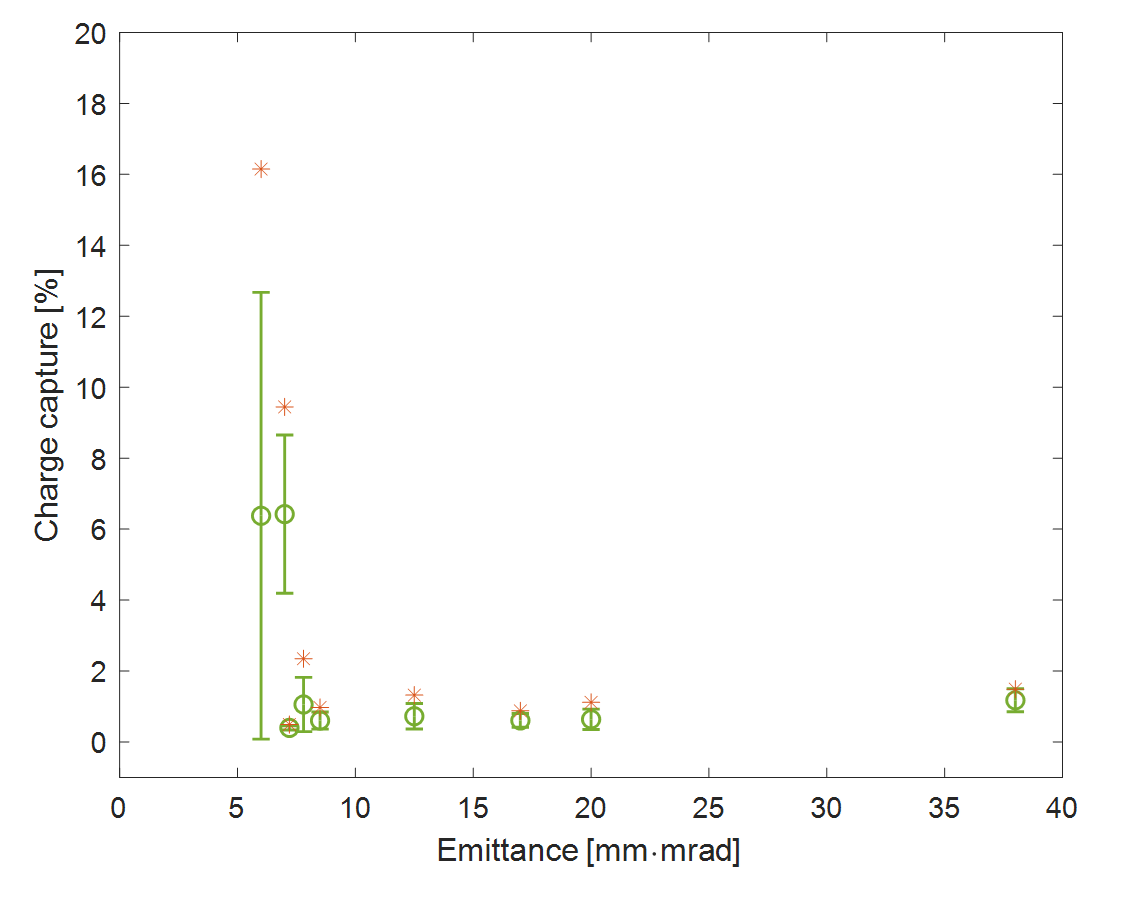}% Here is how to import EPS art
\caption{\label{fig:4} Measured charge capture rate for various input electron beam emittances. Green circles show the average charge capture rate (error bars are the standard deviation) and the orange stars show the maximum recorded capture rates.}.
\end{figure}

A series of experimental runs were carried out with a range of cathode illumination parameters corresponding to the values depicted in Fig. \ref{fig:3}. Due to the electron beam pointing jitter and random overlap with both the plasma column and the proton bunch, a rather variable outcome was observed during the acceleration experiments. Figure \ref{fig:4} shows for example the results of 120 consecutive runs varying the electron beam characteristics. For low emittance electron beams, acceleration events were observed in less than 10\% of the runs, whereas for large emittance this number raised to more than 80\%.

A rather obvious conclusion from this test was that electron beams with low emittance can spatially overlap better with the wakefield accelerator capture window. This optimized spatial matching directly provides a higher charge capture rate. However, it is worth noting that the overall accelerated charge in this situation was essentially equivalent to that measured for high electron beam emittances, which was in the range of 10~–~15~pC. The deviation in charge capture rate was also decreasing as the emittance was increased, owing to the larger electron beam size at the injection point (and consequent immunity to spatial jitter). Overall, the average capture rate obtained was in the range of 2~-~3\% for successful shots. The results, however, portray the possibility of higher charge capture rates employing this acceleration scheme, with a maximum experimentally measured capture rate of more than 16\% (corresponding to 40~pC of accelerated charge) for the best shot. Note that here the driving phenomena for the capture efficiency is not electron beam initial emittance but rather the resulting beam size and spatial overlap with the wakefields at the injection point.

The spatial jitter of the electron beam at the injection point was of several 100s of microns in rms, which allowed us to effectively ‘map’ the wakefield accelerator spatial acceptance window experimentally.For this we relayed on precise BPM readouts averaged over 1 second
prior to each proton shot to estimate the crossing point of
the electron beam into the plasma employing the techniques described in \cite{PhysRevAccelBeams.23.032803}. Because of the complexity of the vapor source, it was not possible to install any beam position or beam size diagnostics close to or along the plasma. Therefore, the last direct measurement of the electron beam was given by a scintillating screen positioned 0.8~m upstream from the entrance of the vapor source. Furthermore, during the acceleration experiment, no screen can be inserted in the beam line, because this would completely absorb the electron beam. This makes the alignment process for the injection extremely challenging due to the uncertainty on the electron transverse beam size at the injection point and due to the different effects of external magnetic fields on the two beam trajectories, given by the very different rigidity. 

The rms transverse size $\sigma_e$ at the crossing point is one of the factors that contributes to the charge capture efficiency. Measuring the size near the crossing point is therefore important. Moreover, including the effect of Earth’s magnetic field on the low-energy beam is crucial to precisely predict the electron beam trajectory only using information provided by BPMs. For detailed information regarding the position and size measurements of the electron beam see \cite{PhysRevAccelBeams.23.032803}. In our experiments, the range of electron beam sizes varied from $\sigma_e$~=~0.19~–~0.55~mm.

To effectively inject the witness bunch into the wakefields, its transverse size must be comparable to the transverse extent of the plasma wakefields. This is given by the plasma skin depth $r_d$:

\begin{equation}
r_d = \frac{c}{\sqrt{n_e e^2/\epsilon_0 m_e}},
\end{equation}

where $n_e$ is the plasma electron density, $e$ the elementary charge, $\epsilon_0$ the vacuum permittivity and $m_e$ the electron mass. For the range of plasma electron densities used in our experiments, the skin depth corresponded to values of just below 0.5 mm.

\begin{figure}[ht]
\includegraphics[width=\linewidth]{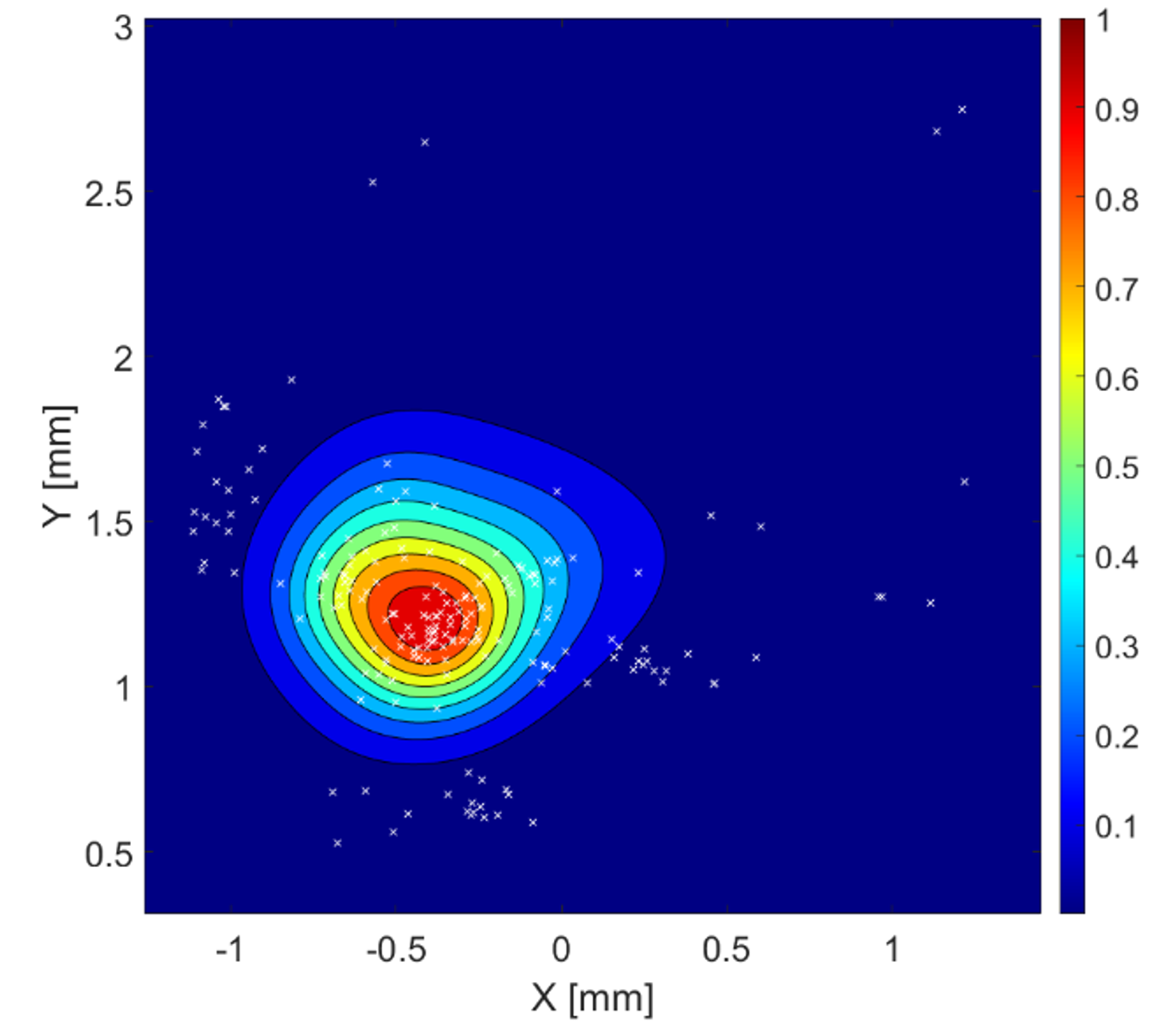}% Here is how to import EPS art
\caption{\label{fig:5} Results of the spatial convolution of the input electron beam with the spatial acceptance window of the wakefields (amplitude was scaled with the average charge capture rate and normalized). The white crosses show all $X_e$, $Y_e$ electron bunch single-shot locations.}.
\end{figure}

 With the extracted values of the electron beam transverse size $\sigma_e$ and the relative position $X_e$, $Y_e$ at the injection point, and the measured accelerated charge, we can now ‘map’ the overall spatial charge capture rate by including enough spatially distributed shots. The resulting plot is a convolution of the different electron beam transverse distributions with the spatial acceptance window of the wakefields, with an amplitude proportional to the electron capture rate per unit area. The results of this ‘map’ are shown in Fig. \ref{fig:5}. As it can be observed, the convolved width was approximately 0.3 mm in both $X$ and $Y$ axes.

Using a similar technique is possible to roughly estimate the acceptance angle of the wakefield accelerator. A detailed description of the estimation of the trajectory of the electron beam in the plasma accelerator can be found in \cite{PhysRevAccelBeams.23.032803}. Here we used that method to also ‘map’ the angles of incidence instead of the positions of the electron beam for each case. The results showed that this window is approximately of 1 mrad in the horizontal direction, although more experiments are needed to produce a complete angle map including the both vertical and horizontal directions. Nevertheless, the results are in good agreement with those predicted by the models described in \cite{CALDWELL20163}.

\section{Conclusions}

The AWAKE experiment at CERN seeks the demonstration of efficient and useful acceleration of electrons in the wake created by a proton beam passing through plasma. These wakefields are used to accelerate the electrons with GV/m strength fields. The region of good trapping is quite large in the space of injection parameters, as compared to the electron beam characteristics. We studied the acceptable electron beam position and angle jitter that enables consistent electron acceleration and compared it with the theoretically calculated wakefield acceptance map. We observed that under full overlap conditions between electron beam and wakefield acceptance window, more than 80\% of the events showed accelerated electrons at the GeV level.

Additionally, we have experimentally measured the spatial wakefield acceptance map that enables high charge capture rate by varying the electron beam characteristics in terms of size, position, trajectory, and emittance. We found that, under optimized conditions, the charge capture rate is larger than 15\% ($\sim$40~pC of accelerated charge for a 385~pC injected electron bunch), therefore approaching the theoretical limit of ~40\%. 

\bibliography{apssamp}% Produces the bibliography via BibTeX.

\end{document}